\documentclass[preprint]{aastex}

\shorttitle{Dwarf Nova HS0907+1902}
\shortauthors{Thorstensen}

\begin{document}

\title{Spectroscopy and Component Masses of the 
Eclipsing Dwarf Nova HS0907+1902}

\author{John R. Thorstensen}
\affil{Department of Physics and Astronomy, Dartmouth College,
Hanover, NH 03755}

\email{john.thorstensen@dartmouth.edu}

\slugcomment{{\it Publications of the Astronomical Society of the Pacific},
in press}

\begin{abstract}
HS0907+1902 was recently discovered to be one of a handful of deeply eclipsing 
dwarf novae with periods longward of the 2 -- 3 hr `gap'.  This paper
presents orbit-resolved spectra and time series photometry of 
an eclipse.  The apparent velocity amplitude of
the M-dwarf secondary is $K_2 = 297 \pm 15$ km s$^{-1}$.  The 
phase of the radial velocities of the H$\alpha$ emission line wings 
agrees accurately (for once) with the phase of the white-dwarf motion deduced
from the eclipse, and an estimate of the 
emission-line velocity amplitude yields 
$K_1 = 115 \pm 7$ km s$^{-1}$.  
The eclipse width is $\Delta \phi = 0.060 \pm 0.005$.  
At face value, these measurements yield mass estimates of
$M_1 = 0.99 \pm 0.12\ M_{\odot}$ for the white dwarf and 
$M_2 = 0.38 \pm 0.06\ M_{\odot}$ for the secondary. The eclipse
width and nominal mass ratio constrain the binary inclination $i$
to be $77^{\circ}.3 \pm 0^{\circ}.9$.  The influence of systematic
uncertainties on these values is discussed; the conclusion that the
white dwarf is somewhat more massive than typical field white
dwarfs appears to be robust.  

The H$\alpha$ emission line profile out of eclipse is only
slightly double-peaked, but the line shows a strong rotational 
disturbance in eclipse.  Models of the line profile 
through eclipse using a flat, Keplerian disk do not
give a good quantitative match to the observations.

\end{abstract}


\section{Introduction}

Cataclysmic variable stars (CVs) are close binaries in which a white dwarf
accretes material from a late-type main-sequence companion.
The average rate at which matter is transferred can vary widely,
leading to a variety of interesting behaviors.  The dwarf novae,
a subset of the CVs, display bright outbursts generally separated
by relatively quiescent intervals; the instability leading to the 
outburst is thought to occur when a critical amount of material 
is accumulated in an accretion disk about the white dwarf, the outburst
itself being caused by a rapid increase in the mass-transfer rate through
the disk.  The orbital periods of dwarf novae range from $\sim 80$ min
up to $\sim 12$ h; there is a significant scarcity in the period
range 2 -- 3 h, loosely referred to as the `gap'.  Dwarf novae
with periods shortward of the gap have distinct behaviors
(dubbed superhumps and superoutbursts) which mark them as SU UMa
stars; dwarf novae significantly longward of the gap generally 
do not show this suite of behaviors, and are called 
U Gem, SS Cyg, or Z Cam stars
(the lattermost exhibiting frequent `standstills' intermediate between
their quiescent and outburst brightnesses).  \citet{war95} gives
a comprehensive review of CVs.

Although a great deal of effort has been devoted to determining the 
masses of the component stars in CVs, systematic effects are
very troublesome.  The apparent motion of the disk's
emission lines can seriously misrepresent the white dwarf motion,
mostly because the disk's Keplerian rotation broadens the line
to a width much greater than the small excursions caused by the 
white dwarf's orbit.  As the observer views the system from different
angles, any slight azimuthal asymmetry in the emission-line
surface brightness creates a velocity modulation which 
masquerades as orbital motion.  These effects are especially
egregious in the class of novalike variables known as SW Sextantis
stars \citep{tho91,dhi97}.  Many of these show
eclipses which reveal the true phase of conjunction, yet the observed
emission line velocities can disagree with the known phase of the white
dwarf motion by as much as $\sim 0.2$ cycle.  Even in the dwarf
novae, significant phase offsets and the presence of `$S$-waves'
arising from the stream-disk collision often complicate the extraction of
dynamical information from the emission lines \citep{mar88}.  Absorption
line velocities of the secondary stars are more
useful, since their interpretation is more straightforward and 
since CV mass ratios $q = M_2/M_1$ are generally
$< 1$, leading to greater velocity amplitude of the secondary.
Even here, interpretation of the velocities can be problematical, since the 
lines arise in an externally-illuminated, non-spherical star with nonuniform
gravity over its surface.  This leads
to a `$K$-correction' which can be difficult to estimate
\citep{wad88}.

Even with these caveats, the discovery of a new eclipsing CV
is cause for celebration.  Eclipses constrain the problem 
sufficiently that careful interpretation of good data
can potentially give reliable measurements of the system parameters.

HS0907+1902, the subject of this paper, was discovered in the
Hamburg Schmidt objective-prism survey \citep{hag95} 
and flagged as a CV candidate based on its spectrum and
a cross-identification with a ROSAT X-ray source 
\citep{bad98}.
It underwent a dwarf nova outburst early in 2000.
\citet{gaen00} show eclipses which reveal a binary
period of 4.2 h, and a minimum-light spectrum with a noticeable
contribution from an M-dwarf secondary.  Because only a handful
of eclipsing dwarf novae are known longward of the gap,
I studied the object further in 2000 April, during
quiescence.

\section{Observations}

The data were all obtained at the MDM Observatory
2.4 m Hiltner telescope, on Kitt Peak.

\subsection{Photometry}

I observed an eclipse on 2000 April 5 UT ($E$ = 330 in 
the ephemeris of \citet{gaen00})
using a SITe 2048$^2$ CCD and a $V$ filter.
To expedite readout the data were binned on chip to 
$2 \times 2$ pixels, and a region of $512 \times 512$ (binned) 
pixels was read.  Each binned pixel subtended $0''.550$.
All exposures were 30 s; with readout and chip preparation,
the cycle time averaged 58 s.  Standardization was not attempted.
I measured four stars
in each image using the IRAF {\tt apphot} task, and took the measurements
differentially with respect to a brighter star east and slightly
south of HS0909+1902.  For this star the USNO A2.0 catalogue
\citep{mon96} gives $\alpha = 9^{\rm h}\ 09^{\rm m}\ 57^{\rm s}.793,$
$\delta = +18^{\circ}\ 49'\ 03.07''$, (ICRS $\approx$ J2000),
and magnitudes $r = 13.3$ and $b = 14.3$.  
Provisionally adopting $V = 13.8$ for this comparison star yields the 
differential light curve shown Figure 1.
The differential magnitudes of a check star similar in brightness 
to the target have a standard deviation of 0.01 mag. 
The adopted magnitude scale yields 
$V \approx 17.6$ in mid-eclipse, in very good agreement with that measured
by \citet{gaen00} with respect to a Tycho-2 star (which lies outside
the present field of view).

The eclipse center occurred at HJD 2451639.7228(5), slightly
earlier than the \citet{gaen00} ephemeris predicts, so
the period can be refined slightly to 0.175444(2) d.  
The eclipse width, measured halfway down the steep sides of the 
profile, is $\Delta \phi = 0.060(5)$.  At the points measured the 
brightest part of the disk is being covered and uncovered,
so $\Delta \phi$ should apply to the disk center's eclipse.
The object is brighter before
eclipse than after, an effect usually attributed to emission
from the stream-disk collision at the `bright spot'. 

The direct images used for the photometry can also be compared to 
the USNO A2.0 catalog \citep{mon96}
to constrain the proper motion.  The first-epoch Palomar Observatory
Sky Survey  plates of this region, on which
the USNO is based, were obtained 1950 Mar.\ 21.  For several
pictures, linear fits were derived between the 
CCD image centers of 6 to 10 stars (excluding HS0907+1902) 
and the corresponding 
USNO A2.0 coordinates (excluding HS0907+1902).  
These yielded transformations 
with root-mean-square errors of $\sim 0''.3$, with excellent 
reproducibility from frame to frame, implying that the
USNO centers dominate the uncertainties as expected.
Applying these transformations to the modern images of HS0907+1902
gave a position 326 mas W and 438 mas S of the USNO
position.  The implied proper motion is 
$11 \pm 6$ mas y$^{-1}$, which is formally insignificant but
suggestive.   

\subsection{Spectroscopy}

On 2000 April 8 UT I obtained spectra covering a single orbit,
and on 2000 April 10 I obtained 
additional exposures of another eclipse.  
The MDM modular spectrograph and SITe 2048$^2$ CCD gave $\sim 3.5$ \AA\ 
resolution from 4200 to 7500 \AA , with severe vignetting toward the 
ends of the range.  Exposures away from 
eclipse were generally 240 s, with shorter exposures when 
eclipses were predicted.  I took wavelength calibration exposures
about once per hour, and the measured
standard deviation of the $\lambda 5577$ night-sky feature's
apparent radial velocity was 7 km s$^{-1}$ over the entire run.  
Flux calibration stars were observed, but the spectrograph
suffers from a mysterious problem which leads to spurious variations in 
the continuum shapes.  With the amount of data obtained these
should largely average out.  Occasional cirrus and seeing variations
at the $1''$ entrance slit further degraded the absolute flux.  

The average spectrum appears nearly identical similar to that
shown by \citet{gaen00}.   The strong, broad emission lines 
typical of dwarf novae at minimum light (detailed in Table 1) 
are accompanied by the absorption features of an M-type secondary.  
The emission lines are only slightly double-peaked, in contrast to the
strongly double-peaked profiles in the 3.8-h eclipsing dwarf nova 
IP Peg \citep{mar88}.  In the averaged spectrum the central 
parts of H$\alpha$ are well-fit by a single Gaussian with a
full-width at half maximum (FWHM) equivalent to 1325 km s$^{-1}$.

To refine the spectral type of the secondary I subtracted a grid 
of scaled spectra of M dwarfs (classified by \cite{boe76})
from the average spectrum and 
examined the results; the best cancellation of the M-dwarf features 
(shown in Figure 2) occurred at M3, in
agreement with G\"ansicke et al.'s estimate of M3.0 $\pm$ 1.5.  
M2- and M4-type spectra did not give satisfactory results, so 
the present estimate is slightly more discriminating 
than theirs.  

Figure 3 shows a single-trailed representation of a portion of the 
spectrum, prepared using the phase-averaging procedure described
by \citet{tay98}.  Because the flux scale varied somewhat
erratically from spectrum to spectrum, the data were rectified
before averaging, which suppresses the eclipse.  The emission 
lines appear brighter in eclipse, which in this representation 
shows that they are eclipsed less deeply than the continuum.  
H$\alpha$ shows a strong rotational disturbance, in contrast
to the lines in some eclipsing novalikes (e.g., \citet{dhi97}).  
Outside of 
eclipse, there is a conspicuous absence of strong double peaks
or $S$-waves; the variations near the line core as a function
of phase appear rather disorganized.  The
M-dwarf absorption features are also visible, most dramatically
the molecular band head near 6160 \AA\ and the atomic lines
just shortward of this.  As expected, their contrast increases
during the eclipse.  The M-dwarf features can be 
traced through the entire orbital cycle, though they may be
less pronounced around phase 0.5 (superior conjunction of the M-dwarf).
They show easily discernible orbital motion.

To measure the M-dwarf motion, I first
co-added the target spectra into 16 phase bins.  
Then, using the {\tt rvsao} package \citep{kur98} I
prepared a velocity-compensated template by summing the spectra of
four M dwarfs with known radial velocities \citep{mar87}.  
A cross-correlation of the region 5900 \AA\ $< \lambda <$ 6500 \AA\
yielded usable radial velocities for 13 of the phase bins.   
The missing points are near M-dwarf superior conjunction where
the features appear weakest.  The cross-correlation routine
gave typical uncertainties of 20 km s$^{-1}$ for the usable
velocities.

To measure the H$\alpha$ emission radial velocities I used convolution
methods outlined by \citet{sch80} and elaborated by \citet{sha83};
briefly, an antisymmetric function consisting of 
positive and negative Gaussians of adjustable separation is 
convolved with the observed line, and the zero of the convolution
(when the contributions of the two sides is equal) is taken as the
line center.  The phase-binned spectra (excluding eclipse) were used for these
measurements; velocities measured from the original spectra
gave very similar results.  A preliminary sine fit to the velocity
variation showed that the phase accurately tracked the
phase of the white dwarf motion as deduced from the eclipse.
Encouraged by this I systematically varied the separation of the
Gaussians in the convolution function in the procedure outlined
by \citet{sha83}, which yielded the diagnostic diagram shown 
in Figure 4.  The aim of this procedure is to find the maximum
separation for which the signal-to-noise is adequate, the 
rationale being that the wings of the line originate toward the 
center of the disk, where the disk should approach azimuthal symmetry.
A separation of 2300 km s$^{-1}$ (full width) 
appeared to give the most reliable results; Figure 5 shows the 
average H$\alpha$ line profile with the adopted measurement function
superposed.  In Figure 6 the emission and absorption velocities are plotted 
together, with sine fits (Table 2) superposed.   

Figure 7 shows H$\alpha$ radial velocities in the vicinity of the eclipse.
There is a very large, regular velocity excursion, first 
toward the red and then toward the blue.  This is the 
signature of a rotating object.  The rotational disturbance is also 
evident in the single-trailed representation (Figure 3). 
A detailed model of the rotational disturbance is described 
in Section 3.2.  A rotational disturbance is seen in the 
eclipsing dwarf nova IP Peg \citep{mar90}.

\section{Interpretation}

\subsection{Masses and Related Quantities}

Let us begin by estimating the component masses, taking
the observed parameters at face value.
We have then
$$q = {M_2 \over M_1} = {K_1 \over K_2} = 0.387 \pm 0.031.$$
This can be combined with the geometry of the Roche-lobe-filling 
secondary and the eclipse width $\Delta \phi$ to yield a 
binary inclination  of $77^{\circ}.3 \pm 0^{\circ}.9$.
With $K_1 + K2 = 412 \pm 16$ km s$^{-1}$, we have from 
Kepler's third law
$$M_1 + M_2 = 1.37 \pm 0.17\ M_{\sun}.$$
The mass ratio then gives
$$M_1  = 0.99 \pm 0.12\ M_{\sun} \ \ \hbox{and}\ \ 
M_2 = 0.38 \pm 0.06\ M_{\sun}.$$

This simple calculation can be criticized on several grounds, and 
the rest of this section is devoted to refining these estimates
and examining how robust they are under various assumptions.

First, the M-dwarf velocity amplitude probably requires 
adjustment because of uneven distribution of the absorption 
features over its surface.  The lower contrast of the absorption
features opposite primary eclipse, and the difficulty of 
obtaining velocities over that range of phases, suggest
that this effect is operating here (though a secondary
eclipse of the M-dwarf by the disk may be contributing also).  
Unfortunately, the present data do not have sufficient photometric 
integrity to accurately estimate the absolute strength of the 
absorption around the orbit, precluding strategies such
as those of \citet{wad88} for estimating the correction. 
Also, the absorption velocities are not precise enough to 
show the deviations from a sinusoid expected for
non-uniform contributions from a Roche lobe \citep{mart89}.
I therefore used a model calculation to crudely estimate the effect.
I first created a model Roche lobe, with 500 segments in an 
icosahedral geodesic pattern similar to that implemented
by \citet{hen92}.
I then synthesized line profiles around the orbit; 
this calculation 
was purely kinematic and geometrical, the contribution of
each surface element to the line profile
being proportional only to the solid angle it subtended at earth.
Because the uncorrected data likely
yield too large a $K_2$, and hence too small a $q$, I 
chose $M_1 = 0.85\ M_{\sun}$ and $M_2 = 0.37\ M_{\sun}$ for the
model, and fixed $i$ at $77^{\circ}.3$.  For these parameters,
the line-of-sight velocity amplitude of the secondary's
center of mass is $K_2 = 276$ km s$^{-1}$.
When I assumed that the entire stellar surface 
contributed equally to the absorption, a sinusoidal fit to the
synthetic velocity curve yielded $K_2 = 276$ km s$^{-1}$, 
indistinguishable from the velocity of the secondary's
center of mass.  With the extreme
assumption that the absorption arises {\it entirely}
from portions of the star not in view of the (pointlike)
white dwarf, $K_2$ increased by 43 km s$^{-1}$.
As a frank guess, I take the actual correction to be
$1/3 \pm 1/6$ of the correction indicated by this simulation,
or $-14 \pm 7$ km s$^{-1}$.
Applying this correction to the actual data changes
$K_2$ to $283 \pm 17$ km s$^{-1}$.  The correction
derived by \citet{wad88} for their velocities of the Z Cha
secondary is practically the same as that 
employed here.  With the revised $K_2$,  repeating the calculations above
yields $q = 0.41 \pm 0.04$, $i = 77^{\circ}.0 \pm 0^{\circ}.9$,
$M_1 = 0.82 \pm 0.14\ M_{\sun}$, and $M_2 = 0.33 \pm 0.07\ 
M_{\sun}$.

The emission-line $K$-velocity presents a more intractable issue, 
since it could misrepresent the 
white dwarf motion while mimicking the expected phase.
The line profiles do not appear strongly double-peaked,
as they do in IP Peg \citep{mar90}, so employing them in the solution 
puts us in the position of using something we don't entirely
understand physically\footnote{Not that this usually 
stops astronomers!}.
On the other hand, the single-trailed spectrum does not
show any obvious $S$-wave (generally the signature of emission
from the stream-disk impact point); detection of 
such a component would create extra complications, at
least.  

Even if we disregard the emission line amplitude entirely,
we can still estimate the white dwarf mass, because the 
condition that the secondary fill its Roche critical lobe
sets the mean density of the secondary as a function of 
orbital period \citep{fau72}.  The spectral type of the 
secondary in HS0907+1902 is typical for its orbital period 
\citep{gaen00,beu98}, which doesn't prove anything but at
least does not indicate trouble.

The density constraint is often implemented by assuming 
a main-sequence mass-radius relationship for the
secondary.  This approach was used by  
\citet{pat84} to derive a semi-empirical relation between
orbital period and $M_2$, which predicts 0.40 $M_{\sun}$ 
in this case \citep{gaen00}; encouragingly, this is in 
reasonable agreement with the mass estimated earlier.
Fixing $i = 77^{\circ}$, the corrected $K_2$ then requires 
$M_1 = 0.92\ M_{\sun}$.  
If we allow a generous uncertainty of $\pm 0.2\ M_{\sun}$
for $M_2$, this propagates to $\pm 0.2\ M_{\sun}$ in
$M_1$. The $\pm 17$ km s$^{-1}$ uncertainty in $K_2$
propagates to $\pm 0.1\ M_{\sun}$, and even a rather wide
3-degree uncertainty in $i$ propagates to only
$\pm 0.02\ M_{\sun}$ in $M_1$.  Summed in quadrature these
yield $$M_1 = 0.92 \pm 0.22\ M_{\sun}.$$

This estimate is rather crude; a somewhat more sophisticated
one recognizes that 
the main-sequence mass-radius relation is uncertain, and that
CV secondaries may not be main-sequence stars.
\citet{bar00} model CV evolution and find that different
(plausible) assumptions predict substantially different
values of $M_2$ at a given orbital period.  
In addition, 
\citet{wad90} gives a formalism for the uncertainty in
$M_1$ when the Roche-filling constraint is employed and when
various parameters have been measured; the present 
calculation is his Case 3 
($i$ measured, $K_2$ measured, $K_1$ presumed unknown).
He parameterizes the uncertain secondary star mass-radius 
relation as $R_2 = c_2 M_2$.  The range of evolutionary models 
computed by \citet{bar00} (their Table 2) has 
$0.96 \le c_2 \le 1.88$ at
this period, with $0.18\ M_{\sun} \le M_2 \le 0.51\ M_{\sun}$.
If we take the spread of the model parameters as a guide to the
plausible range, we can
characterize this as $M_2 = 0.33 \pm 0.15\ M_{\sun}$ and 
$c_2 = 1.4 \pm 0.4$.  Taking the 
other parameters and their uncertainties as above, and using
the \citet{wad90} formalism for the error, gives
$$M_1 = 0.86 \pm 0.18\ M_{\sun}.$$

\citet{ber92} find that 
field DA white dwarfs have masses strongly clustered near
0.56 $M_{\sun}$; the white dwarf in 
HS0907+1902 appears to be significantly more massive 
than this.  

\subsection{The Eclipse of H$\alpha$}

The rotational disturbance in the H$\alpha$ line presents
an opportunity to probe the accretion disk's structure. 
To explore this I constructed synthetic line profiles
for a model eclipsed disk, and attempted to match them
to the observations.  The model disk had 600 segments, divided into 
20 logarithmically-spaced annuli and 30 azimuthal zones.
Ingress and egress phases for each segment were calculated
using a method similar to that employed by \citet{hor85};
the minimum potential along the ingress or egress 
line of sight generally matched the critical Roche potential
to better than 1 part in 10$^{4}$. 
The velocity of each segment was approximated as circular
Keplerian motion around the white dwarf combined with the 
systemic orbital rotation, and the emissivity was 
taken to be a power law in distance from the white dwarf,
$I \propto r^{-\beta}$ .  
The line radiation from each segment was assumed to be
broadened in wavelength using a Gaussian profile with an 
adjustable $\Delta \lambda$  (full
width half maximum, or $\sim 2.35 \sigma_{\lambda}$), constant among the 
different segments; even if there is no intrinsic broadening, $\Delta \lambda$
can account for the instrumental resolution.
The inner and outer disk radii $r_{\rm min}$ and $r_{\rm max}$ 
were adjustable and expressed as a fraction of the distance from the
white dwarf to L1.  The system parameters were fixed at representative 
values ($M_1 = 0.90\ M_{\sun}$, $M_2 = 0.4\ M_{\sun}$, and 
$i = 77^{\circ}.3$).  I ran models with $0.05 \le r_{\rm min} \le 0.3$,
$0.4 \le r_{\rm max} \le 0.7$, $0 \le \beta \le 2$, and 
$5 \le \Delta \lambda \le 20$.   The largest value of 
$\Delta \lambda$  corresponds to 390 km s$^{-1}$ of
broadening at H$\alpha$.  It was chosen in order to smear
out the double peaks predicted by all simple disk models.
Line profiles were computed for each set of parameters at
binary phases 0.85 through 0.15, and at each simulated
phase a radial velocity was computed using the same double-Gaussian
convolution algorithm as for the real data.

As expected, the models qualitatively reproduced the observed
behavior, but quantitative agreement was poor.  The solid
curve in Fig.\ 7 is among the simulations which most closely
resembled the data, and the fit is not particularly good.  
Comparison of the models with the data yielded the following
conclusions.  When the inner disk radius was set small
($0.05 \le r_{\rm min} \le 0.2$), the peak-to-peak
amplitude of the rotational disturbance was much larger
than observed.   Unless the outer disk radius was made quite
large (0.6 to 0.7), the slope of the velocities 
across mid-eclipse was too steep.  But when the outer disk
radius was adjusted to match the slope at mid-eclipse,
the maximum and minimum velocities occurred rather too
far from eclipse, as in the example shown.  Rather surprisingly,
the broadening $\Delta \lambda$ and disk index $\beta$ did not affect
these conclusions significantly, but avoiding strongly
double-peaked profiles out of eclipse required 
$\Delta \lambda \ge 15$ \AA .  All of the models showed
conspicuously double-peaked profiles in a small interval 
around mid-eclipse, a feature not seen in the data.

The poor match to the data suggests that the H$\alpha$ emission
in HS0907+1902 does not entirely arise in a flat, Keplerian disk.
A rotating, but sub-Keplerian, disk wind may contribute.  
It is also possible that the system parameters are 
off (particularly the inclination), or that the disk 
is significantly flared.  The remedy adopted in the model for the 
lack of strong double peaks -- a very large 
smoothing in $\lambda$ -- is quite {\it ad hoc};  a
non-Keplerian component might fill in the line center in a
more natural way.  

\section{Conclusion}

HS0909+1902 presents a splendid opportunity for determining the 
masses of the component stars in a dwarf nova.  The secondary
star's radial velocities are easily measured, and the 
emission-line velocities appear to be unusually well-behaved.
The particular masses found here will undoubtedly be superseded
as better data are acquired.  Spectra with finer
velocity resolution and better spectrophotometric integrity
should yield a very reliable value for the M dwarf's 
center-of-mass motion, and detailed tomographic and eclipse 
studies of the emission lines should test the assumption that  
$K_1$ is a trustworthy measure of the white dwarf motion.
Although the present work will not be the last word, it helps
spread the good news.  The conclusion that the white
dwarf mass is larger than typical field white dwarf
masses does appear fairly robust, as it is can be deduced
using the absorption-line velocities alone, without reference
to the more devious emission-line velocities.

The behavior of the H$\alpha$ emission line through eclipse
presents a puzzle.  While the expected rotational disturbance
is observed, simple models do not give a good quantitative match.
At first this is somewhat disappointing, but it opens an 
avenue for further research.  With improved system parameters,
the region of space occulted by the secondary as a function
of time will be even better constrained.  Because a rotational
disturbance {\it does} occur, we know that {\it something} is 
rotating in the occulted region. With physically-motivated 
parameterizations of various possible effects 
(sub-Keplerian disk winds, Stark broadening, and flared disks
to name some possibilities), it may be possible to 
assemble a physically-based model of the velocity field
in the emitting region.  Spectra with improved signal-to-noise,
phase redundancy, time resolution, and (especially) photometric 
integrity will be key to this effort.

\acknowledgments

I'd like to thank Joe Patterson for urging me to observe this
object, the MDM staff for expert observatory 
support, and the referee, Richard Wade, for timeliness,
care, and useful suggestions.   I am grateful for the 
support of the National Science Foundation through 
grant AST-9987334.

\clearpage

\begin{figure}
\plotone{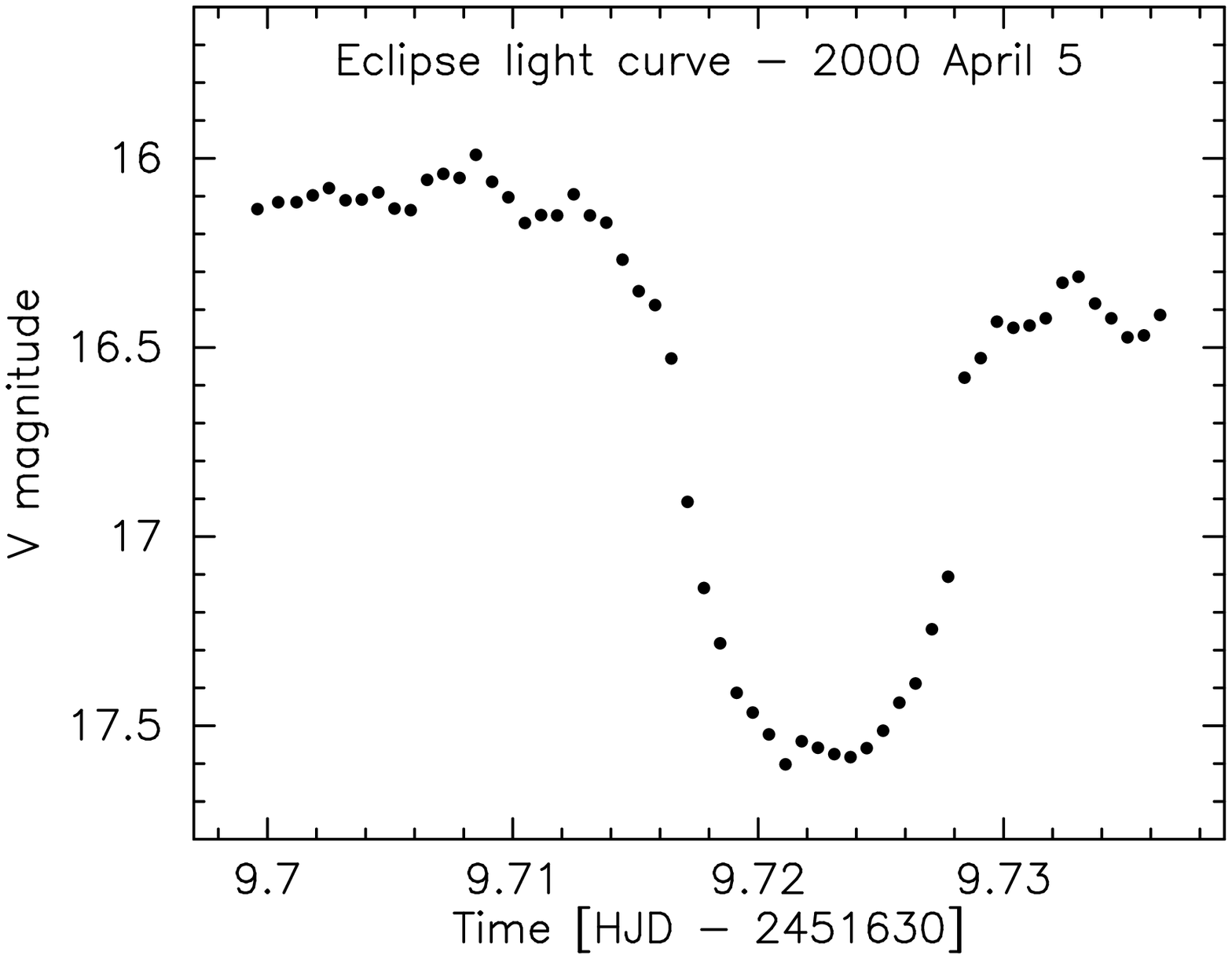}
\figcaption[lghtcrv.eps]{Differential $V$ light curve of an eclipse
of HS0907+1902.  The comparison star (see text) is assumed to have 
$V = 13.8$.  \label{fig1}}
\end{figure}
\clearpage

\begin{figure}
\plotone{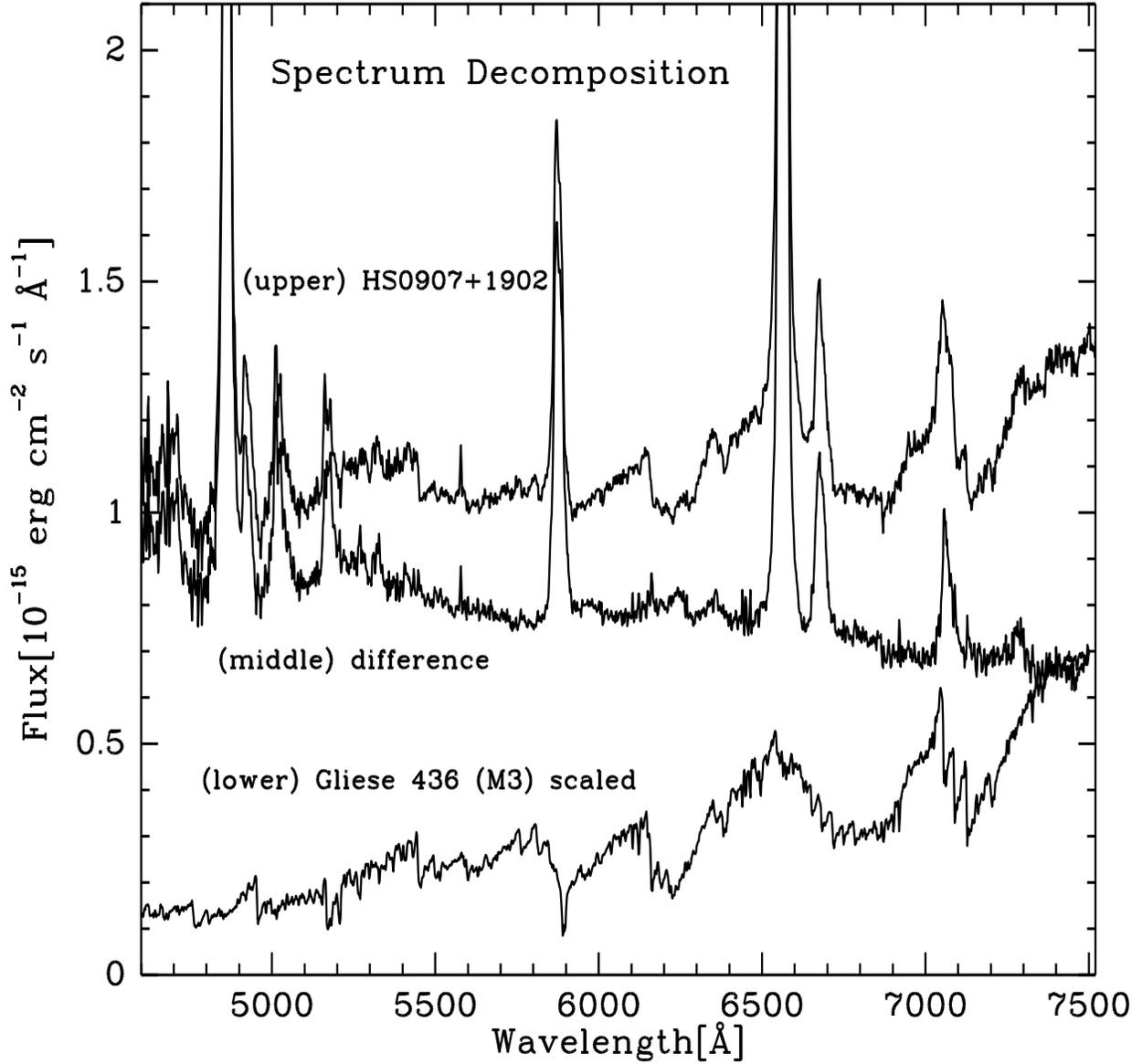}
\figcaption[decomp.eps]{{\it upper trace:} the average spectrum
of HS0907+1902; {\it lower:} the M3-type dwarf Gliese 436
scaled to $V \sim 18.0$; {\it middle:} the difference
of these two. \label{fig2}}
\end{figure}
\clearpage

\begin{figure}
\epsscale{0.9}
\plotone{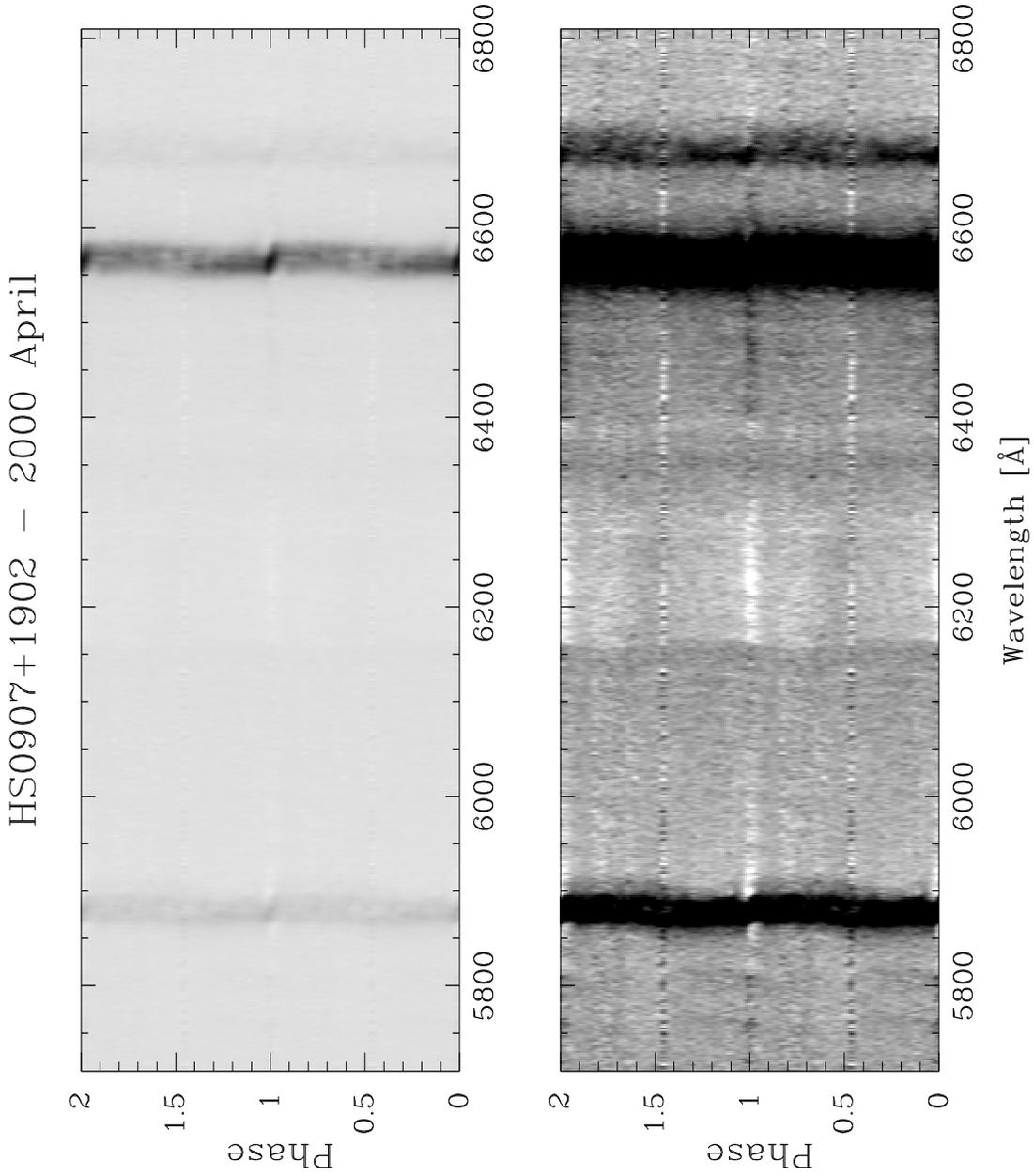}
\figcaption[singletrail.eps]{Greyscale plot of the spectra,
arranged by binary phase and displayed as a single-trailed spectrum .   
All data are shown twice for phase continuity.  The individual
spectra were rectified before the figure was assembled.
The scaling of the lower panel enhances
visibility of weaker features (e.g. the M-dwarf absorption).
The upper panel shows the same data scaled to show the
emission-line cores.  The horizontal feature near $\phi = 0.48$
is an artifact of a short interruption in the data stream.
\label{fig3}}
\end{figure}
\clearpage

\begin{figure}
\epsscale{0.7}
\plotone{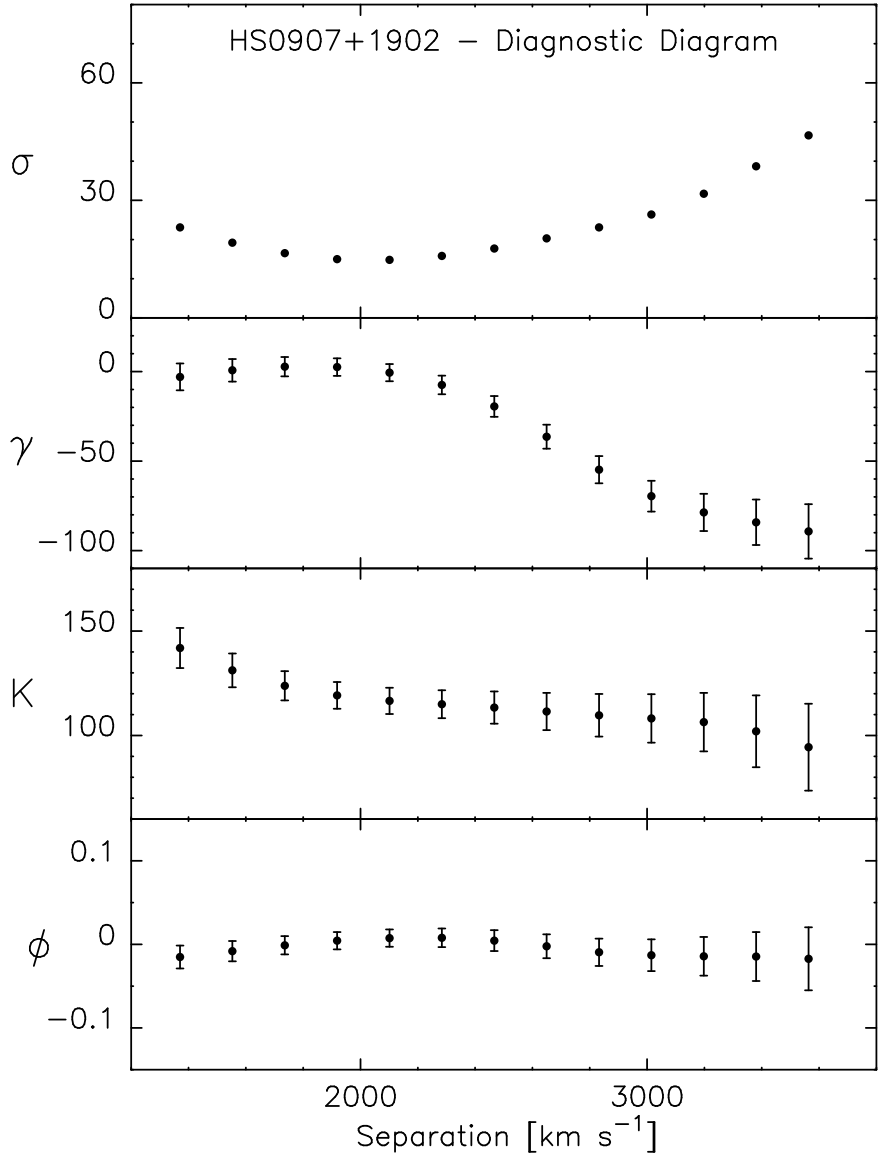}
\figcaption[diagnost.ps]{Diagnostic diagram for the H$\alpha$
emission line velocity measures.  The fit parameters are defined
in the notes to Table 2.
\label{fig4}}
\end{figure}
\clearpage

\begin{figure}
\epsscale{0.9}
\plotone{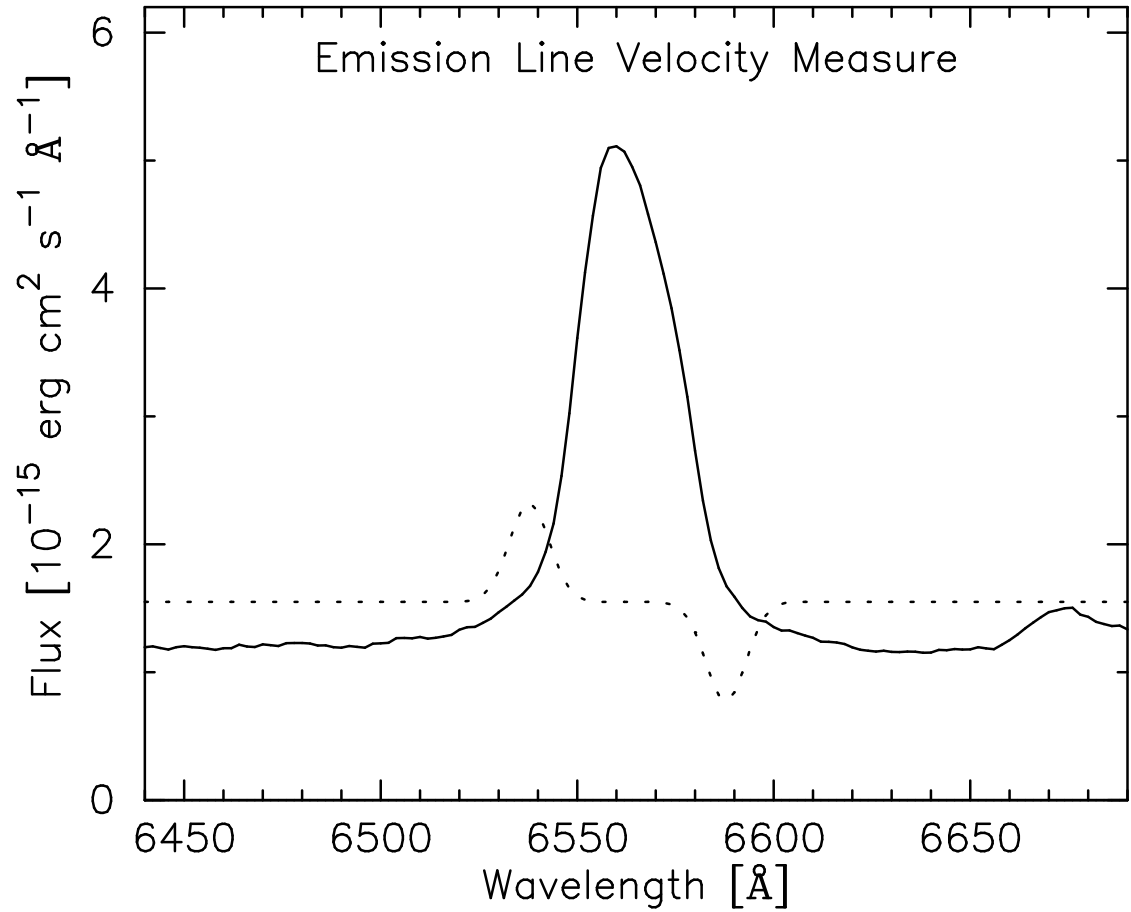}
\figcaption[linemeas.ps]{The H$\alpha$ profile from the average
spectrum (solid curve) together with the adopted antisymmetric 
function used for the line measurement (dot-dash curve).  The 
vertical scale shown applies to the line profile, and the
scale for the convolution function is not shown.  Its background
level is zero.
\label{fig5}}
\end{figure}
\clearpage

\begin{figure}
\epsscale{0.7}
\plotone{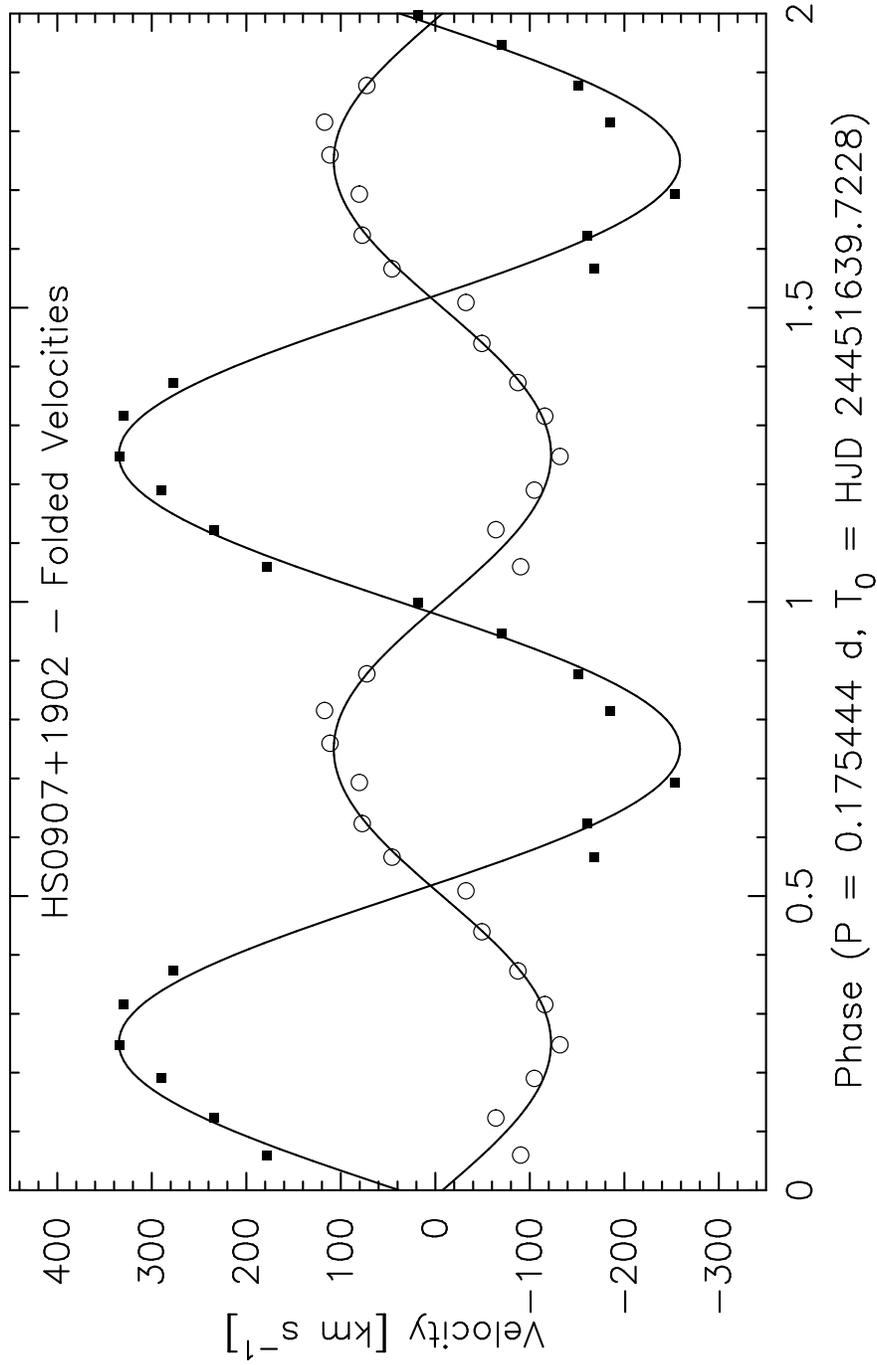}
\figcaption[vcurves.ps]{Velocities of the H-alpha emission (open
circles) and M-dwarf absorption (solid circles) measured from the 
phase-binned spectra.  Best-fit sinusoids are superposed.  For
continuity, all data are repeated for a second cycle.
\label{fig6}}
\end{figure}
\clearpage

\begin{figure}
\epsscale{0.7}
\plotone{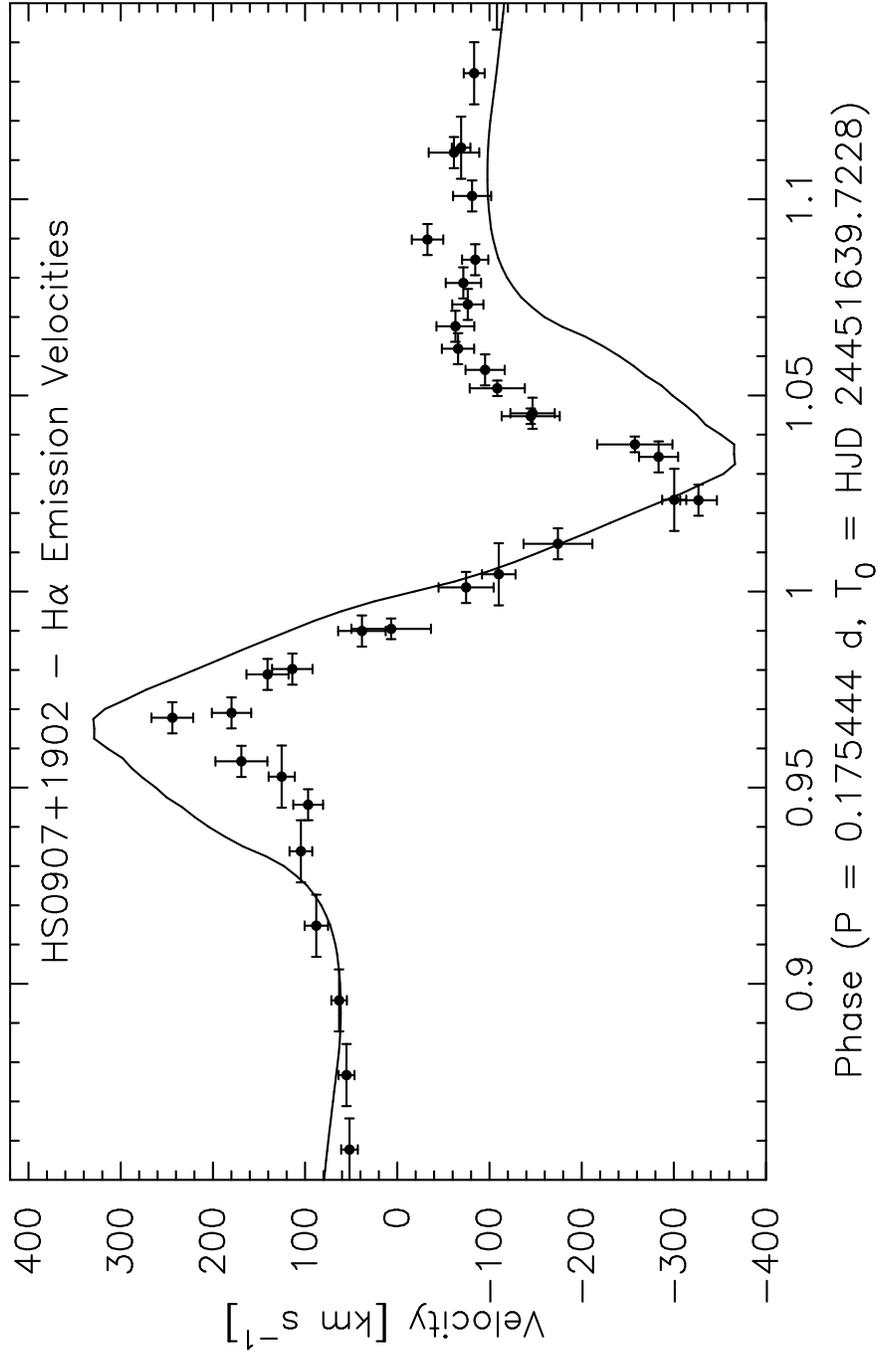}
\figcaption[eclipservmatch.ps]{Radial velocities of H$\alpha$ derived from
individual exposures near eclipse.  The horizontal error bars
show the duration of the exposures, and the vertical error 
bars are counting-statistics errors propagated through the measurement
procedure.  The solid curve is derived from the eclipse of a flat,
Keplerian model accretion disk described in the text, with 
$r_{\rm min} = 0.3$,$r_{\rm max} = 0.7$,
$\beta = 2$, and $\Delta \lambda = 15$ \AA .}
\end{figure}
\clearpage





\begin{deluxetable}{lcc}
\tablecaption{Emission Line Measures \label{tbl-1}}
\tablewidth{0pt}
\tablehead{
\colhead{Line} & 
\colhead{Equivalent Width\tablenotemark{a}}  & 
\colhead{Flux} \\
& \colhead{(\AA )} & \colhead{(10$^{-14}$ erg cm$^2$ s$^{-1}$)}
}
\startdata
H$\gamma$ & 51 & 6.6 \\
HeI 4471 & 15: & 1.6: \\
HeII 4686 & 7: & 0.6: \\ 
H$\beta$ & 110 & 8.8 \\
HeI 4921 & 12 & 1.0 \\
HeI 5015 & 16 & 1.3 \\
FeI:: 5169 & 11 & 0.9 \\
HeI 5876 & 38 & 2.9 \\
H$\alpha$ & 161 & 12.3 \\
HeI 6678 & 17 & 1.3 \\
HeI 7067 & 16 & 1.1 \\
\enddata

\tablenotetext{a}{Equivalent width measured after subtraction of
secondary star contribution.}


\end{deluxetable}

\clearpage

\begin{deluxetable}{lccccc}
\tablecaption{Sinusoidal fits \label{tbl-2}}
\tablewidth{0pt}
\tablehead{
\colhead{Data set} & 
\colhead{$\Delta \phi$ \tablenotemark{a}}  & 
\colhead{$K$ \tablenotemark{b}} &
\colhead{$\gamma$} & 
\colhead{$\sigma$ \tablenotemark{c}} & 
\colhead{$N$}\\ 
& & \colhead{(km s$^{-1}$)} &
\colhead{(km s$^{-1}$)} &
\colhead{(km s$^{-1}$)} &  
}
\startdata
H$\alpha$ emission & $+0.013 \pm 0.008$  & $116 \pm 5$  & $-18 \pm 4$ & 19 & 42\\
H$\alpha$ emission (binned) & $+0.008 \pm 0.011$ & $115 \pm 7$ & $-7 \pm 6$ 
&16 & 14\\
Absorption (binned) & $-0.001 \pm 0.003$ & $297 \pm 15$ & $38 \pm 11$ & 33 & 13\\
\enddata

\tablenotetext{a}{Phase offset based on the local
eclipse ephemeris, HJD $24451639.7228 +
0.175444 E$, where $E$ is an integer.  
The values here are offsets from the phases
expected for the white dwarf (emission) and 
red dwarf (absorption) motions; $\Delta \phi$
is positive for events which occur late.
}

\tablenotetext{b}{$K$ and $\gamma$ are respectively
the amplitude and mean of the fitted sinusoid.}

\tablenotetext{c}{Uncertainty of a single point 
derived from the scatter around the best fit.}


\end{deluxetable}

\clearpage

\end{document}